\documentclass[10pt,twocolumn,preprintnumbers,superscriptaddress,nofootinbib,aps,prl]{revtex4}

\usepackage{graphicx}
\usepackage{amsmath}
\usepackage{srcltx}
\usepackage[utf8]{inputenc}
\usepackage[colorlinks]{hyperref}
\usepackage{times}
\usepackage{subfigure}
\newcommand{\vev}[1]{\langle {#1} \rangle}
\newcommand{\lsim}{\lesssim}
\newcommand{\gsim}{\gtrsim}

\newcommand{\eq}[1]{Eq.~(\ref{#1})}

\newcommand{\ord}[1]{\mathcal{O}{(#1)}}
\newcommand{\beq}{\begin{equation}}
\newcommand{\eeq}{\end{equation}}

\newcommand{\qcond}{\vev{\bar q \, q}}
\newcommand{\scond}{\vev{\bar s \, s}}

\newcommand{\mP}{M_{\rm P}}

\newcommand{\appropto}{\mathrel{\vcenter{
  \offinterlineskip\halign{\hfil$##$\cr
    \propto\cr\noalign{\kern2pt}\sim\cr\noalign{\kern-2pt}}}}}

\begin{document}

\pagestyle{plain}

\title{\boldmath Exploring Strange Origin of Dirac Neutrino Masses at Hadron Colliders}

\author{Hooman Davoudiasl\footnote{email: hooman@bnl.gov}
}

\affiliation{High Energy Theory Group, Physics Department, Brookhaven National Laboratory,
Upton, New York 11973, USA}

\author{Ian M. Lewis\footnote{email: ian.lewis@ku.edu}
}

\affiliation{Department of Physics and Astronomy, University of Kansas, 
Lawrence, Kansas, 66045 USA}

\author{Matthew Sullivan\footnote{email: msullivan1@bnl.gov
}
}

\affiliation{High Energy Theory Group, Physics Department, Brookhaven National Laboratory,
Upton, New York 11973, USA}


\begin{abstract}

We consider the possibility that Dirac neutrino masses may be a manifestation of chiral symmetry breaking via 
non-perturbative QCD dynamics.  The key role played by light quarks in this mechanism 
can naturally lead to signals that are accessible to hadron colliders.  Bounds  
from charged meson decays imply a dominant effect from the strange quark condensate.  We propose a model for Dirac neutrino mass generation with an extra Higgs doublet at the TeV scale and significant coupling to strange quarks and leptons.  Current data on $D-\bar D$ mixing constrain the allowed parameter space of the model, and  
a 100 TeV $pp$ collider would either discover or largely exclude it.  A distinct feature of this scenario is that measurements of the of charged Higgs leptonic branching ratios can distinguish between ``normal" and ``inverted" neutrino mass hierarchies, complementing future determinations at neutrino oscillation experiments.

\end{abstract}
\maketitle


\section{Introduction}

The origin of non-zero but tiny neutrino masses $m_\nu \lsim 1$~eV remains an open fundamental question in particle physics.  The smallness of 
$m_\nu$ implies very suppressed interactions with the Higgs and its vacuum expectation value $\vev{H}=v/\sqrt{2} \approx 174$~GeV.  Hence, 
typically, neutrino mass models either involve very massive fermions which are inaccessible to direct observations, or else negligible Yukawa 
couplings.  These considerations generally leave only indirect rare processes, such as neutrinoless double $\beta$ decay, as the possible signals 
of the underlying neutrino mass generation mechanism.  

Given that $m_\nu \neq 0$ requires a connection with a source of electroweak symmetry breaking (EWSB), the involvement of $\vev{H}$ seems like  a necessary ingredient in any plausible model.  However, there is another source of EWSB in the Standard Model (SM), the light quark condensate $\qcond \sim - (300 ~\text{MeV})^3$, with $q\in \{u,d,s\}$.  This effect -- generated by non-perturbative QCD interactions -- has been largely ignored as a possible origin of neutrino masses.  The much smaller contribution from $\qcond$ to EWSB suggests that it could have far less suppressed interactions with neutrinos in an underlying mass generation mechanism.  Therefore, one may generically expect that key states in such a mechanism would be accessible to hadron colliders, where light quarks are important initial states for producing new particles.

In this work, we will consider generation of Dirac neutrino masses through QCD quark condensation.  First, we will examine the general aspects of this setup using effective field theory (EFT).  We will find that constraints from meson decays lead us to consider the strange quark condensate $\scond$ as the dominant source of neutrino masses in this scenario.  In order to realize the mechanism, we will introduce a model that includes a TeV scale Higgs doublet $H_2$ with $\ord{1}$ couplings to strange quarks as well as  right-handed partners assumed for each SM neutrino flavor.  In our convention, the SM-like Higgs doublet $H$ is denoted by $H_1$. We will show that, depending on the specific values of couplings, heavy scalars can be resonantly produced and detected in di-jet or ``lepton + missing energy" final states, under minimal assumptions regarding the model.  Prospects for flavor experiments and potential consequences for supernova observations will also be briefly discussed.  Prior works that have considered neutrino mass generation from QCD dynamics include Refs.~\cite{Thomas:1992hf,McDonald:1996cs,Ibanez:2001nd,Davoudiasl:2005ai,Babic:2019zqu}. 

\section{EFT Framework}

Let us begin with an effective theory description and consider the following dimension-6 interaction 
\beq
O_D = \zeta\frac{[\bar Q\,s]\epsilon[\bar L \, \nu_R]}{M_D^2} + {\small \rm H.C.}\,,
\label{OD}
\eeq
where $Q$ and $s$ are the second generation quark doublet and the strange quark singlet, respectively, $L$ is a lepton doublet and $\epsilon$ is the 2-dimensional Levi-Civita symbol contracting the $SU(2)_L$ indices.  The right-handed field $\nu_R$ is the Dirac partner of the left-handed $\nu_L$ component of $L$.  In \eq{OD}, $\zeta$ is a $3\times 3$ matrix corresponding to three generations of $L$ and $\nu_R$, assuming three massive neutrinos.  Note that $[\bar Q\,s]$ has the electroweak quantum numbers of an anti-Higgs doublet and condensation of its iso-spin $-1/2$ component can generate a neutrino mass term.  We have suppressed all generational indices in \eq{OD}.  The effective scale $M_D$ is related to mass scales in an ultraviolet (UV) completion and generally includes the effect of couplings in such a theory. In principle, one could also have a similar operator with $s$ replaced by the right-handed up or down quark.  However, as will be discussed below, those choices will lead to significant constraints.  To obtain a neutrino mass $m_\nu \sim 0.1$~eV upon quark condensation, we then need $M_D\sim 16$~TeV, where we have used $\scond \approx -(300~\text{MeV})^3$ \cite{Davies:2018hmw}. 

Constraints from charged meson $P^+$ decays of the type $P^+ \to e + \text{``missing energy"}$ were  considered in Ref.~\cite{Davoudiasl:2005ai}.  We adapt the results of Ref.~\cite{Davoudiasl:2005ai} for the partial width for meson decay $P^+ \to e^+\, \nu_R$
\beq
\Gamma(P^+ \to e^+ \, \bar \nu_R) = \frac{\sum_i |\zeta_{ei}|^2}{64 \pi\, M_D^4} f_P^2 \mu_P^2 m_P\,,
\label{Gamma}
\eeq
where the sum over $i$ is over right-handed neutrinos, $f_P$ and $m_P$ are the decay constant and mass of $P$.  
The parameter $\mu_P$ is given by 
\beq
\frac{m_\pi^2}{2 \bar m}\quad;\quad \frac{m_K^2}{m_s + \bar m}\quad;\quad \frac{m_{D_s}^2}{m_{c} + m_s}\,,
\label{muP}
\eeq
for $P=\pi, K$, or $D_s$.  The mass of the $D_s^+$ is $m_{D_s}=2.0$~GeV.  In the above, $\bar m = (m_u + m_d)/2$; $m_u$, $m_d$, $m_s$, and $m_c$ are the up, down, strange, and charm quark masses, respectively.  The current relevant branching ratio limits are~\cite{Zyla:2020zbs}
\begin{eqnarray}
&{\rm Br}(\pi^+\rightarrow e^+\,\nu)&=\,(1.23\pm0.004)\times 10^{-4}\\
&{\rm Br}(K^+\rightarrow e^+\,\nu)&=\,(1.582\pm 0.007)\times 10^{-5}\\
&{\rm Br}(D_s^+\rightarrow e^+\,\nu)&<\, 8.3\times 10^{-5}\quad \text{(90\% C.L.).}\label{BRs}
\end{eqnarray}
The decay constants are $f_{\pi}=130$ MeV, $f_{K}=156$ MeV, and $f_{D_s}=250$~MeV; and the lifetimes are $\tau_{\pi^+}=2.6\times10^{-8}$~s, $\tau_{K^+}=1.2\times 10^{-8}$~s, and $\tau_{D_s^+}=5.0\times 10^{-13}$~s~\cite{Zyla:2020zbs}.  The total widths of the mesons can be calculated by $\Gamma_P=1/\tau_P$.

To find bounds, we calculate when the partial width in Eq.~(\ref{Gamma}) saturate the uncertainty bands of the branching ratios\footnote{{\color{black} While the quoted uncertainties on the branching ratios are purely experimental, we have also checked our calculations against the $R$-ratios~\cite{Zyla:2020zbs} where the theory uncertainties~\cite{Cirigliano:2007xi,Bryman:2011zz,Bryman:2021teu} are subdominant.  The conclusions are the same.}}  and assume $\sum_i |\zeta_{ei}|^2\sim1$.  The results then suggest when $P=\pi^+$ the current limits on these decay require $M_D\gsim 60$~TeV~\cite{Davoudiasl:2005ai} and for $P=K^+$ we have $M_D\gsim 100$~TeV, which would not allow large enough neutrino masses, based on the preceding discussion.  Note, however, that if $Q$ is the second generation quark doublet $(c, s)$, then $P^+=D_s^+$ and as we will show later, the current limits on $D_s^+\to e^+ + \text{``missing energy"}$ would allow $M_D\gsim 3$~TeV, which can lead to the observed neutrino masses.  Hence, we will assume the dominance of the second generation interactions in \eq{OD}, which will then require a singlet $s$ to get a quark condensate, since the charm quark $c$ is too heavy to condense.  

Another aspect of phenomenology in our framework is to make sure that right-handed neutrinos do not get populated in the early Universe \cite{McDonald:1996cs} and become extra degrees of freedom during Big Bang Nucleosynthesis (BBN) and cosmic microwave background eras.  This can be arranged for $M_D \gsim 10$~TeV \cite{Davoudiasl:2019lcg,Davoudiasl:2021syn}, by assuming a low reheat temperature $T_{\rm rh}\lsim 0.1$~GeV which nonetheless is high enough to allow a consistent cosmology, given that BBN only requires $T_{\rm rh}\gsim 4$~MeV \cite{Hannestad:2004px} and there is currently no known cosmological input requiring a higher reheat temperature.  

We also point out that supernova constraints from SN1987A can be satisfied if $M_D\gsim 13$~TeV \cite{Thomas:1992hf}, which does not exclude the range of parameters relevant to our model and can be accommodated.  Note that the dominant contribution in our model is from the strange quark, which is presumably not as abundant in a supernova as the up and down quarks making up the nucleons.  Hence, the above supernova constraint -- which was derived assuming neutrino masses from $\vev{\bar u u}$ and the coupling of first generation quarks to right-handed neutrinos -- could be somewhat relaxed.  

Before going further, we would like to make a comment about the possibility of ruling out our model, solely by improving the bound (\ref{BRs}) on ${\rm Br}(D_s^+\rightarrow e^+\,\nu)$.  This bound, which from the above discussion yields $M_D\gsim 3$~TeV, is based on the Belle experiment data set of $913$~fb$^{-1}$ \cite{Belle:2013isi}.  Given that our model requires $M_D\sim 16$~TeV, and optimistically assuming that future measurements are only statistically limited, to probe our model one needs a data set that is larger by $(16/3)^4$, {\it i.e.} $\sim 700$~ab$^{-1}$.  The Belle II experiment is expected to accumulate $50$~ab$^{-1}$ over its entire run \cite{Belle-II:2018jsg}, and is currently the only foreseeable machine that could improve Belle results.  Hence, we do not expect that the model proposed here be ruled out in the foreseeable future, as the constraints on ${\rm Br}(D_s^+\rightarrow e^+\,\nu)$ get more stringent.  This suggests that a 100 TeV collider could be the only envisioned facility that can potentially discover or largely rule out our scenario\footnote{We would like to note that since our model consists of Dirac neutrinos, discovery of neutrinoless double beta decay would rule out the scenario proposed here.}, in the context of a UV model, as will be described later in this work.


\subsection{Flavor Model} 
\label{<sec:flavor>}

Introduction of a new Higgs doublet which couples to light fermions could induce unwanted flavor violating effects.  To mitigate such effects, we will adopt the Spontaneous Flavor Violation (SFV) \cite{Egana-Ugrinovic:2018znw,Egana-Ugrinovic:2019dqu} framework.  This would allow us to have interactions of light fermions with $H_2$ with $\ord{1}$ strength, necessary to obtain sufficiently large neutrino masses, without causing large deviations from flavor constraints.  These couplings, in particular interaction with light quarks, would then allow significant resonant production of $H_2$ at high energy hadron colliders, which can lead to a testable neutrino mass mechanism. We use SFV of the up-type, since the down-type would require large $b$ quark Yukawa coupling $\gsim 1$.

The general 2HDM Yukawa sector with neutrinos would look like 
\begin{eqnarray}
&\displaystyle\sum_{a=1}^2&-\lambda^a_u \bar Q \, \epsilon \, H_a^* \,u - \lambda^a_d \bar Q \, H_a \, d - 
\lambda^a_\nu \bar L \, \epsilon\, H_a^* \,\nu_R - \lambda^a_\ell \bar L \, H_a \,\ell\,\nonumber\\
&&+{\rm H.C.}
\label{eq:Yukawa}
\end{eqnarray}  
In the up-type SFV framework, these Yukawa coupling matrices will take the form
\begin{equation}
\begin{array}{ccc}
\lambda^1_u = V_{CKM}^\dagger Y_u,& \quad \lambda^1_d = Y_d,& \quad  \lambda^1_\ell = Y_\ell  \\
\lambda^2_u = \xi V_{CKM}^\dagger Y_u,& \quad \lambda^2_d = K_d,& \quad \lambda^2_\ell = \xi_\ell Y_\ell ,
\label{eq:uptypeSFV}
\end{array}
\end{equation}
where $Y_u$, $Y_d$, and $Y_\ell$ are the diagonal SM Yukawa couplings in the mass basis for up-type quarks, down-type quarks, and charged leptons, respectively.  Here, $V_{CKM}$ is the CKM matrix, $K_d=diag(\kappa_d,\kappa_s,\kappa_b)$ is flavor diagonal with real entries $\kappa_{d,s,b}$; $\xi$ and $\xi_\ell$ are real constants. For the neutrino Yukawa couplings, we will use the scheme
\begin{eqnarray}
\lambda^1_\nu &=& 0 \nonumber \\
\lambda^2_\nu &=& V_{PMNS} K_\nu ,\label{eq:lamnu}
\end{eqnarray}
where $K_\nu=diag(\kappa_{\nu,1}, \kappa_{\nu,2}, \kappa_{\nu,3})$ with $\kappa_{\nu,i}$ real and, in our model, proportional to the neutrino masses $m_{i}$.  
  Assuming the above structure of Yukawa couplings, we then find that the effective operator in \eq{OD} can be obtained by integrating out $H_2$, with the identifications $\zeta \to \kappa_s \lambda^2_\nu$ and $M_D \to M_{H_2}$, where $M_{H_2}$ is the heavy Higgs doublet mass.   Inserting $\langle s\bar{s}\rangle$ and rotating to the mass basis, \eq{OD} becomes a neutrino mass term and we can make the identification
\begin{eqnarray}
\kappa_{\nu,i}=\frac{M_{H_2}^2}{\kappa_s\,\langle \bar{s}s\rangle}m_i.\label{eq:kapnu}
\end{eqnarray} 
Interestingly, perturbative unitarity gives an upper and lower bound on $\kappa_s,\kappa_{\nu,i}$ from Eq.~(\ref{eq:kapnu}).  Through the usual perturbative unitarity arguments~\cite{Chanowitz:1978mv}, the scattering $f\bar{f}\rightarrow H_2\rightarrow f\bar{f}$ gives an upper bound $|\kappa_s|,|\kappa_{\nu,i}|\leq \sqrt{8\pi}$.  Using Eq.~(\ref{eq:kapnu}), this translates into a lower bound on $\kappa_s$:
\begin{eqnarray}
|\kappa_s|=\frac{M_{H_2}^2}{|\kappa_{\nu,i}|\, |\langle \bar{s}s\rangle|}m_i\geq \frac{M_{H_2}^2}{\sqrt{8\pi}\, |\langle \bar{s}s\rangle|}m_i,
\end{eqnarray}
and similarly for $\kappa_{\nu,i}$.  The most constraining limit comes from the heaviest neutrino, which depends on if the neutrino masses have a normal or inverted hierarchy.  As discussed below, the cosmological bound of $\sum m_i<0.12$~eV~\cite{Ivanov:2019hqk} can place an upper bound of $0.03$~eV and $0.016$~eV on the heaviest neutrino in the normal and inverted hierarchies, respectively.  Hence, normalizing the neutrino mass to $0.03$~eV, we find
\begin{eqnarray}
\sqrt{8\pi}>|\kappa_s|,\,|\kappa_{\nu,i}|>5.5\times 10^{-3}\left(\frac{M_{H_2}}{5~{\rm TeV}}\right)^2\frac{m_i}{0.03~{\rm eV}}.
\end{eqnarray}
Although the lower bound is small, in the scenario presented here, it is not possible to decouple $H_2$ by setting its couplings with fermions to zero.

\begin{figure}[t]
\centering
\includegraphics[width=\columnwidth]{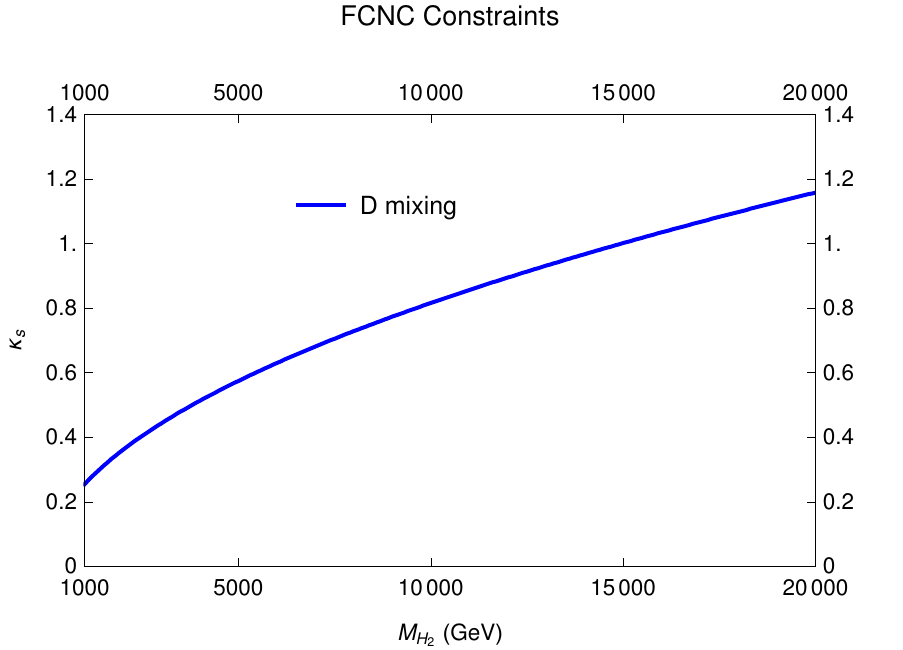}
\caption{Limits on the strange quark coupling $\kappa_s$ as a function of the mass $M_{H_2}$ of the heavy Higgs doublet from $D-\bar{D}$ mixing constraints.}
\label{fig:Dmixing}
\end{figure}

For the up-type parameters, we focus on $\xi\ll 1$ and $\kappa_d, \kappa_b\ll \kappa_s$, with $\kappa_s\lsim 1$. We will justify these parameter values, as well as $\lambda^1_\nu = 0$, within a UV framework later. Although flavor-changing neutral currents are suppressed in the SFV framework, many processes still constrain the allowed parameters. For our strange-dominant scheme, though, the only additional relevant constraint comes from $D-\bar{D}$ mixing. Reference~\cite{Egana-Ugrinovic:2019dqu} calculates the contribution from charged Higgs diagrams in this SFV model to $D-\bar{D}$ mixing. Since the SM contribution is not currently known, we bound the new physics contribution to the mass difference by the 95.5\% upper limit on the observed value from Ref.~\cite{LHCb:2021ykz}. The bounds from this procedure are shown in Fig.~\ref{fig:Dmixing}.

\section{UV Framework}

Next, we will introduce a UV framework for how the assumed Yukawa structures in Eqs.~(\ref{eq:Yukawa}), (\ref{eq:uptypeSFV}), and (\ref{eq:lamnu}) may emerge at the TeV scale.  We have implicitly assumed that the two Higgs doublets $H_1$ and $H_2$ have negligible mixing and hence are, to a good approximation, mass eigenstates.  Here, $\vev{H_1}\neq 0$ is responsible for masses of all fermions except for neutrinos.

We note that, so far, it is not forbidden to have the usual Dirac neutrino Yukawa coupling to the SM Higgs doublet $H_1$, containing the real scalar observed at $m_H\approx 125$~GeV.  In order to eliminate this interaction -- which requires an exponentially small Yukawa coupling $\sim 10^{-12}$ -- in favor of the QCD-mediated mechanism, we assume that both $H_2$ and $\nu_R$ are charged under a global symmetry, which we will take to be a $Z_2$ parity: 
\beq
Z_2(H_2) = Z_2(\nu_R) = -1.
\label{Z2}
\eeq   
Other SM fields are $Z_2$-even.

The above charge assignment, however, does not allow the coupling to the $s$ quark assumed in \eq{eq:Yukawa}, unless it has a chiral $Z_2$-odd parity.  Since $H_1$ is $Z_2$-even, a $Z_2$-odd strange quark would be massless, which is not acceptable. We note that Refs.~\cite{Thomas:1992hf,Davoudiasl:2005ai} had entertained the possibility that the condensate was formed by an up quark, which would remain massless.  This possibility is disfavored by hadron phenomenology and lattice investigations \cite{FermilabLattice:2018est,Alexandrou:2020bkd}.  In this work, we will describe possible model building solutions that will not lead to a massless light quark.  
Hence, the above $s$ quark coupling to $H_2$ must be induced through spontaneous $Z_2$ breaking.  To accomplish this, we postulate a new scalar $\phi$, with $Z_2(\phi)=-1$, interacting through the dim-5 operator,
\beq
c_s \frac{\phi\, H_2 \bar Q s}{\Lambda}\,.
\label{dim5}
\eeq             
Once $\vev{\phi}\neq 0$, it leads to the assumed coupling in \eq{eq:Yukawa}, given by $\lambda^2_s= c_s\vev{\phi}/\Lambda$, with $c_s$ being a complex constant.

Our UV framework also needs to explain why a similar dim-5 operator $\sim \phi \bar L \epsilon H_1^* \nu_R$, leading us back to a Yukawa coupling to the SM Higgs, is not generated.  This could be simply a result of mass hierarchies in the underlying  model.  That is, one could generate the interaction in \eq{dim5}, for example, through the exchange of a $Z_2$-odd vector-like quark $\chi$, of mass $m_\chi \gsim 1$~TeV, that has the SM quantum numbers of a right-handed strange quark $s$.  In the far UV, this requires the following interactions
\beq
y_s \phi \bar \chi \frac{(1 + \gamma_5)}{2} s + y_\chi H_2 \bar Q \frac{(1 + \gamma_5)}{2} \chi + {\small \rm H.C.}\,,
\label{phi-int}                         
\eeq 
which would yield $c_s/\Lambda = y_s y_\chi/m_\chi$.  If there is no analogue $Z_2$-odd vector-like lepton, or if it is very heavy, then we would not obtain a dim-5 operator $\sim \phi \bar L \epsilon H_1^* \nu_R$, or else it would have negligible effect.  

One may also worry that non-perturbative $Z_2$-violating gravitational effects of the type  
\beq
\frac{\phi \bar L \, \epsilon \, H_1^*\, \nu_R}{\mP}\,,
\label{Grav}
\eeq
suppressed by the Planck mass $\mP \sim 10^{19}$~GeV, 
could induce large competing effects. However, for $\vev{\phi}/\mP \ll 10^{-12}$ these effects are expected to be negligible.  In what follows, we will assume $\vev{\phi}\sim$~(1-1000) TeV, which satisfies the above condition.  In \eq{dim5}, it is implicitly assumed that $\chi$ can be integrated out to yield an effective theory description valid for $\vev{\phi} \lsim m_\chi$.  Therefore, if $\vev{\phi}$ is near the lower end of this interval, we could also expect to have TeV-scale $\chi$ particles, which are $SU(3)_c$ triplets and can provide another potential signal at hadron colliders.  Given the above model building considerations, we can account for the features of the assumed Yukawa sector in \eq{eq:Yukawa}.        

\section{Hadron collider phenomenology}
Now we discuss the hadron collider phenomenology of our model.  This model consists of four heavy scalars from $H_2$: charged Higgs bosons $H^\pm$, a scalar $H$, and a pseudoscalar $A$.  The major production modes of these scalar are through the $\kappa_s$ coupling, introduced in the flavor model:
\begin{eqnarray}
s\bar{s}\rightarrow H/A, c\bar{s}\rightarrow H^+,\,{\rm and}\, s\bar{c}\rightarrow H^-.
\end{eqnarray}
The distinctive feature of this model, is that in a wide range of the parameter space the charged Higgs decay into leptons $H^-\rightarrow \ell_i\bar{\nu}_j$ is substantial.  At hadron colliders, the neutrino flavors of the final state cannot be determined and must be summed over.  The partial widths of the charged Higgs bosons into leptons are then
\begin{eqnarray}
\Gamma(H^\pm\rightarrow \ell_i\nu)=\frac{1}{16\,\pi}\left(\lambda^2_\nu {\lambda^2_\nu}^\dagger\right)_{ii}\,M_{H_2}, 
\end{eqnarray}
and the partial width into jets is
\begin{eqnarray}
\Gamma(H^\pm \rightarrow c\,s)&=&\frac{3}{16\,\pi}\kappa_s^2\,M_{H_2}\nonumber\\
&\approx&2.4\times 10^{-3}\,\left(\frac{\kappa_s}{0.2}\right)^2\,M_{H_2}.
\end{eqnarray}

Using the values for the neutrino mixing matrices and mass differences found in Ref.~\cite{Esteban:2020cvm}, as well as the $\lambda_\nu^2$ parameter values above, we find for the normal hierarchy the partial widths into different lepton final states are:
\begin{widetext}
\begin{eqnarray}
\Gamma(H^\pm \rightarrow e\,\nu)&\approx&3.3\times10^{-5}\left[1+12 \left(\frac{m_1}{0.03~{\rm eV}}\right)^2\right]\left(\frac{M_{H_2}}{5~{\rm TeV}}\right)^4\left(\frac{\kappa_s}{0.2}\right)^{-2}\,M_{H_2}\label{eq:norm_el}\\
\Gamma(H^\pm \rightarrow \mu\,\nu)&\approx&6.1\times10^{-4}\left[1+0.63\,\left(\frac{m_1}{0.03~{\rm eV}}\right)^2-3.5\times 10^{-3}\cos\delta_{CP}\right]\left(\frac{M_{H_2}}{5~{\rm TeV}}\right)^4\left(\frac{\kappa_s}{0.2}\right)^{-2}\,M_{H_2}\label{eq:norm_mu}\\
\Gamma(H^\pm \rightarrow \tau\,\nu)&\approx&4.6\times10^{-4}\left[1+0.83\,\left(\frac{m_1}{0.03~{\rm eV}}\right)^2+4.7\times 10^{-3}\cos\delta_{CP}\right]\left(\frac{M_{H_2}}{5~{\rm TeV}}\right)^4\left(\frac{\kappa_s}{0.2}\right)^{-2}\,M_{H_2}\label{eq:norm_tau},
\end{eqnarray}
\end{widetext}
where $m_1$ is the lightest neutrino mass.  The cosmological bound on the sum of neutrino masses is $\sum m_i\lesssim 0.12$~eV~\cite{Ivanov:2019hqk}.  For the normal hierarchy, this translates to a bound on the lightest neutrino mass of $m_1\lesssim 0.030$~eV, to which we have normalized the lightest neutrino mass.  As can be seen immediately, the dependence of $\delta_{CP}$ is minimal.  Hence, we set $\delta_{CP}=0$ for simplicity.

\begin{figure*}[tb]
\begin{center}
\subfigure[]{\includegraphics[width=0.45\textwidth,clip]{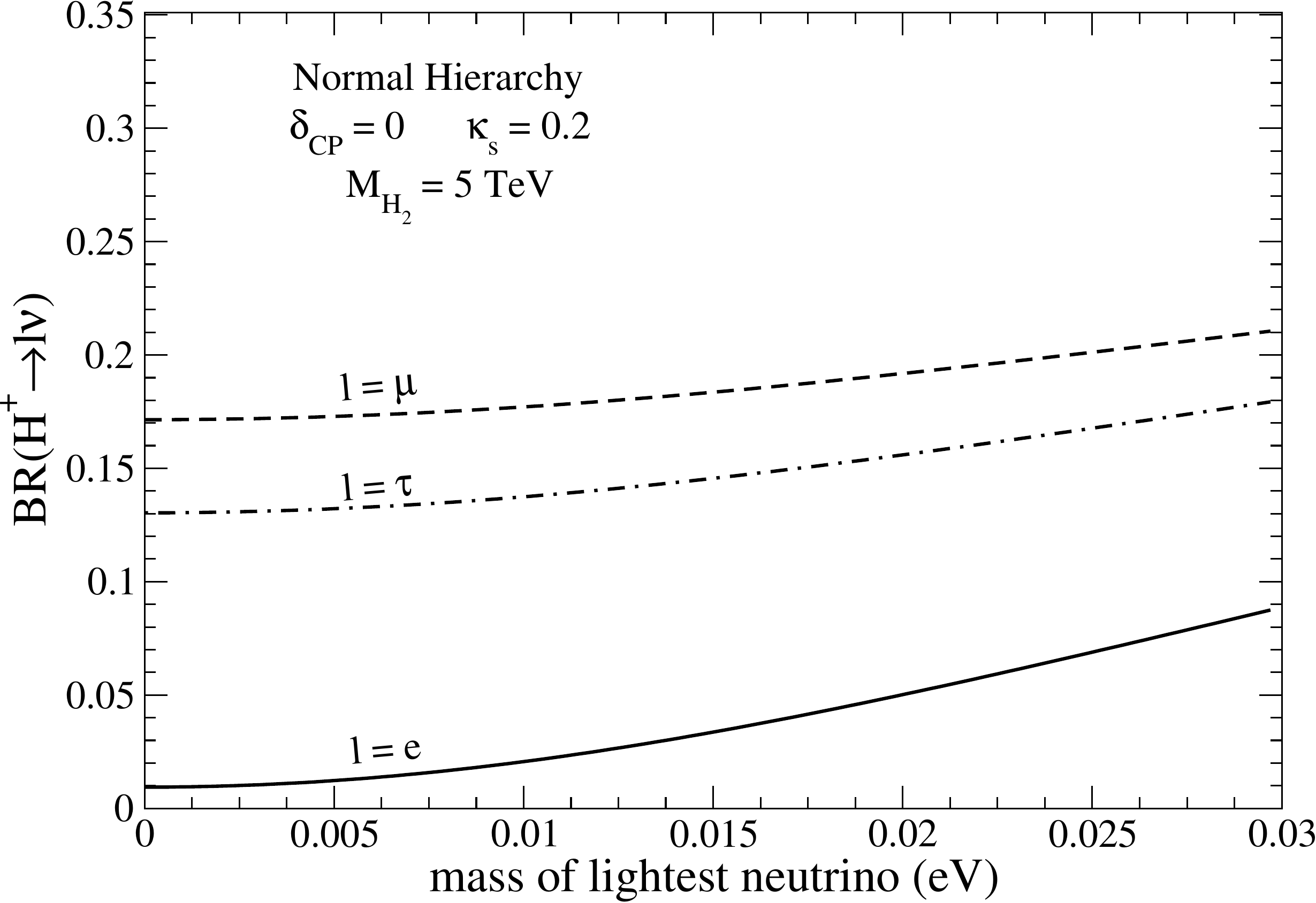}\label{fig:BR_norm}}
\subfigure[]{\includegraphics[width=0.45\textwidth,clip]{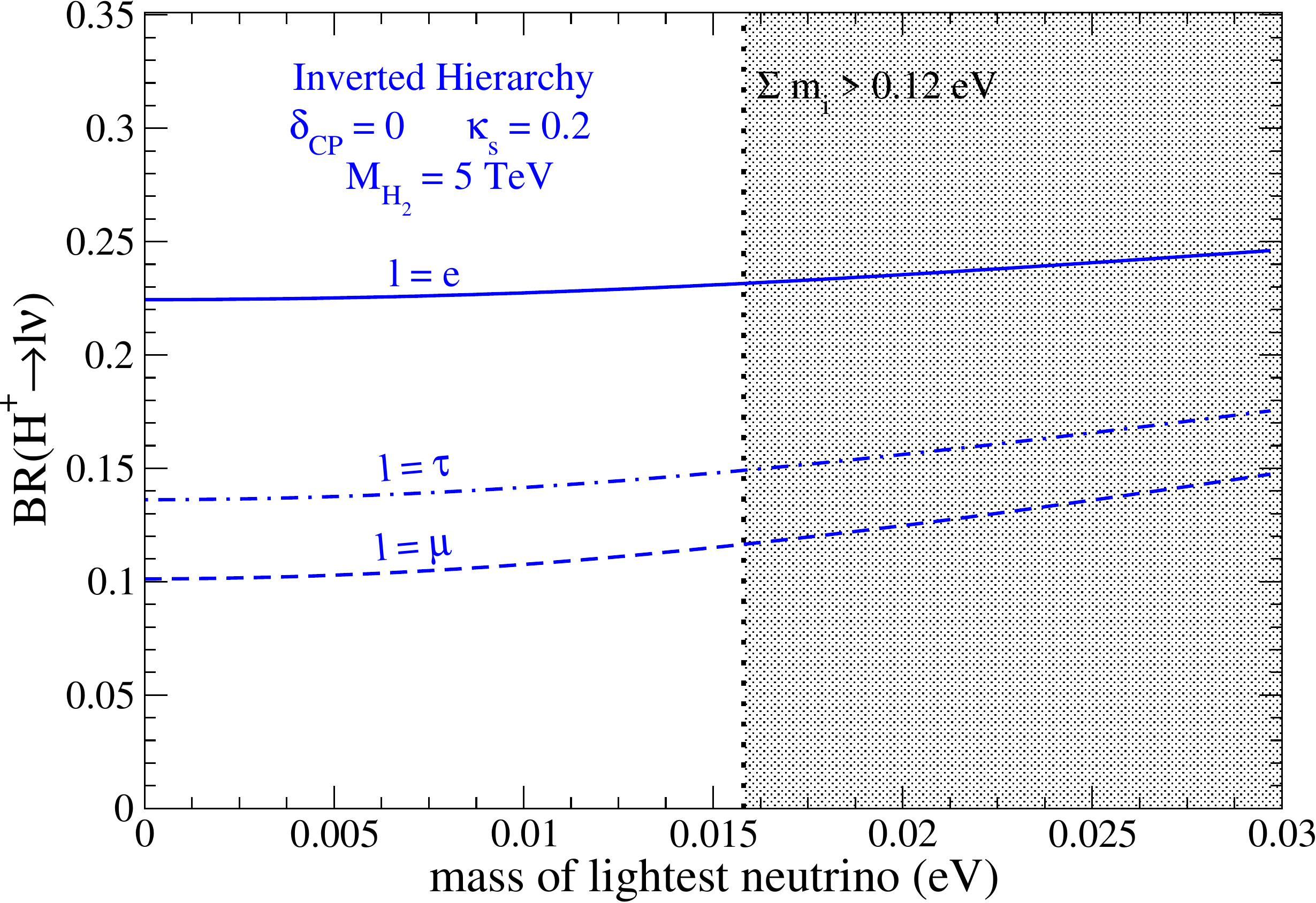}\label{fig:BR_inv}}\\
\subfigure[]{\includegraphics[width=0.45\textwidth,clip]{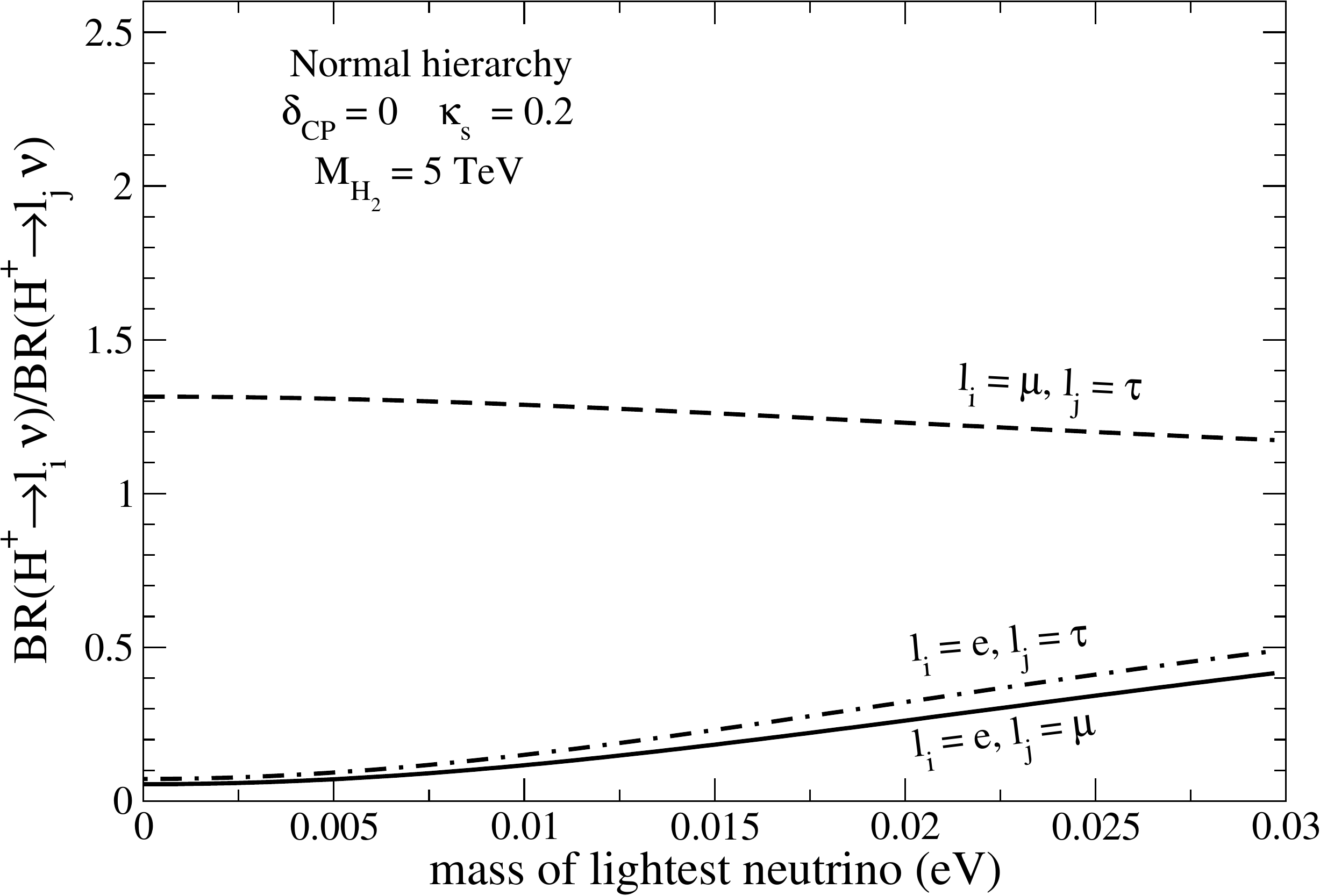}\label{fig:BR_rat_norm}}
\subfigure[]{\includegraphics[width=0.45\textwidth,clip]{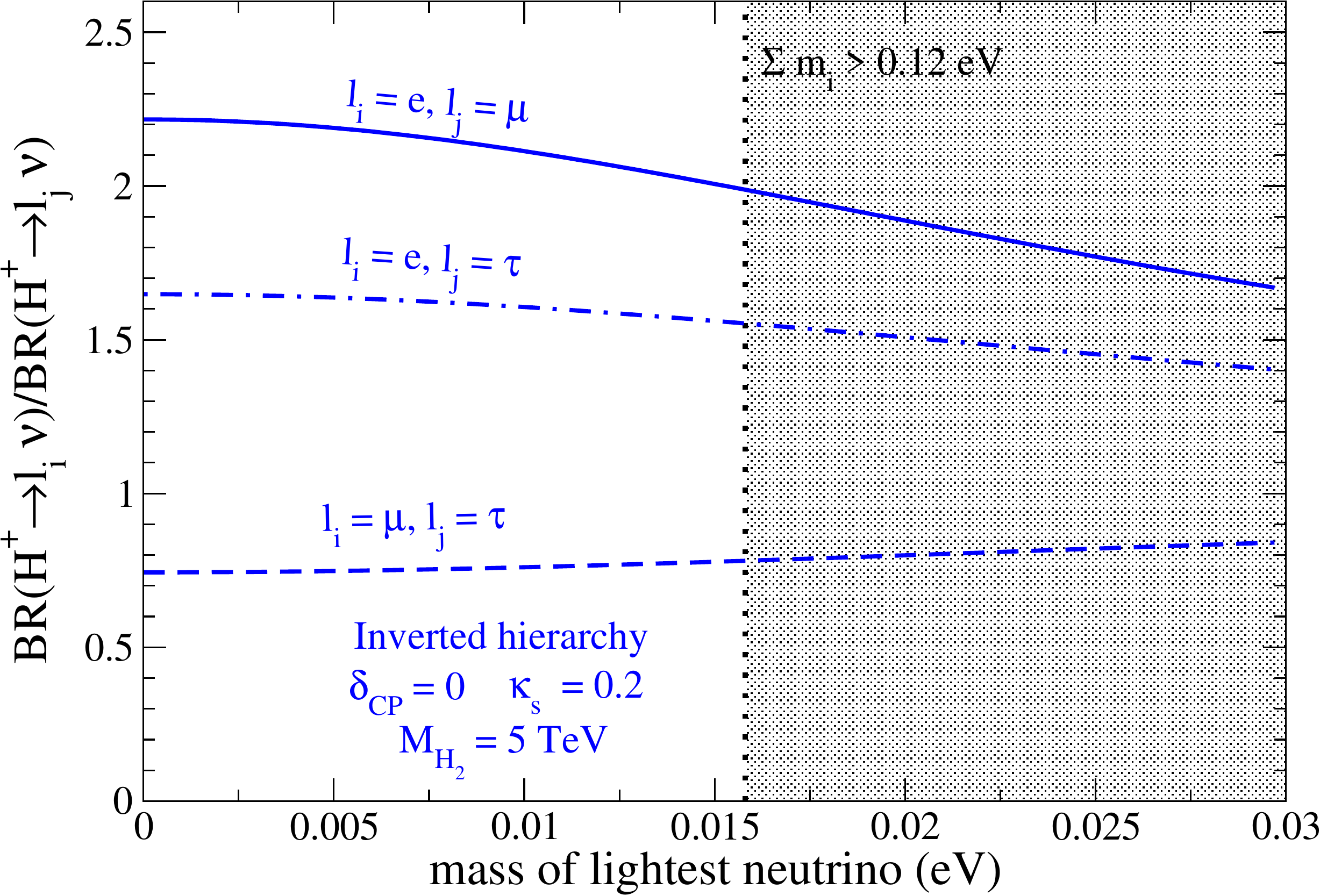}\label{fig:BR_rat_inv}}
\caption{\label{fig:BRs} (a,b) Branching ratios of $H^\pm$ as a function of the lightest neutrino mass into (solid) electrons, (dashed) muons, and (dot-dashed) taus.  (c,d) Ratios of branching ratios between different lepton final states as a function of the lightest neutrino mass: (solid) electron to muon, (dashed) muon to tau, and (dot-dashed) electron to tau.  The normal hierarchy is shown in (a,c) and the inverted hierarchy in (b,d).  The $H_2$ mass is set to 5 TeV, and gray regions correspond to when $\sum m_i > 0.12$~eV.  }
\end{center}
\end{figure*} 

Figures~\ref{fig:BRs}(a,c) shows the (a) branching ratios of $H^\pm$ into leptons and (c) ratios of branching ratios into leptons as a function of hte lightest neutrino mass  in the normal hierarchy.  As is clear in these figures, as well as Eqs.~(\ref{eq:norm_el}-\ref{eq:norm_tau}), the branching ratio into electrons is by far the smallest and, while we expect more $\mu$s than $\tau$s, the branching ratios into muons and taus are comparable. This can be understood by noting that in the normal hierarchy $m_3$ is the heaviest neutrino and couples most strongly to heavy Higgses.  Since $\theta_{13}$ is the smallest mixing angle, the electron couples least strongly.  Since $\theta_{12}$ and $\theta_{23}$ are comparable, the coupling to muons and taus are comparable.  However, as is also clear in the figures and Eqs.~(\ref{eq:norm_el}-\ref{eq:norm_tau}), the branching ratio into electrons is most sensitive to the value of the lightest neutrino mass.

In the inverted hierarchy, the situation is very different.  Now $m_2$ is the heaviest neutrino with $m_1$ closest in mass.  In this case, the partial width into electrons is larger than muons and taus:
\begin{widetext}
\begin{eqnarray}
\Gamma(H^\pm \rightarrow e\,\nu)&\approx&1.0\times10^{-3}\left[1+0.11 \left(\frac{m_3}{0.016~{\rm eV}}\right)^2\right]\left(\frac{M_{H_2}}{5~{\rm TeV}}\right)^4\left(\frac{\kappa_s}{0.2}\right)^{-2}\,M_{H_2}\label{eq:inv_el}\\
\Gamma(H^\pm \rightarrow \mu\,\nu)&\approx&4.6\times10^{-4}\left[1+0.24\,\left(\frac{m_3}{0.016~{\rm eV}}\right)^2-4.7\times 10^{-3}\cos\delta_{CP}\right]\left(\frac{M_{H_2}}{5~{\rm TeV}}\right)^4\left(\frac{\kappa_s}{0.2}\right)^{-2}\,M_{H_2}\label{eq:inv_mu}\\
\Gamma(H^\pm \rightarrow \tau\,\nu)&\approx&6.2\times10^{-4}\left[1+0.18\,\left(\frac{m_3}{0.016~{\rm eV}}\right)^2+3.5\times 10^{-3}\cos\delta_{CP}\right]\left(\frac{M_{H_2}}{5~{\rm TeV}}\right)^4\left(\frac{\kappa_s}{0.2}\right)^{-2}\,M_{H_2}\label{eq:inv_tau},
\end{eqnarray}
\end{widetext}
now $m_3$ is the lightest neutrino mass and the cosmological bound $\sum m_i\lesssim 0.12$ eV requires $m_3\lesssim 0.016$~eV.  Figures~\ref{fig:BRs}(b,d) show the branching ratio results for the inverted hierarchy.  As is clearly seen, in the inverted hierarchy charged Higgs decays into electrons always dominates the lepton modes.  Additionally, the dependence on $\delta_{CP}$ is still very weak, and all modes depend on the exact value of the lightest neutrino mass at a comparable level.

Our model of neutrino masses has distinct predictions for charged Higgs decays that separate the normal and inverted hierarchies.  In the normal hierarchy the charged Higgs decays into muons dominate, followed by taus, and electrons.  In the inverted hierarchy, things are, well, inverted: decays into electrons dominate, then taus, then muons.   Hence, our model has the remarkable feature, that if true, the neutrino hierarchy could be determined by counting the number of electrons and muons coming from charged Higgs decays.

The neutral scalars, $H$ and $A$, can only decay into neutrinos and $s\bar{s}$.  Again, since neutrinos are not detectable, we must sum over them in the final state.  The partial widths are then
\begin{widetext}
\begin{eqnarray}
\Gamma(A/H\rightarrow s\bar{s})&=&\frac{3\,\kappa_s^2}{32\,\pi}M_{H_2}\approx 1.1\times10^{-3}\left(\frac{\kappa_s}{0.2}\right)^2M_{H_2}\\
\Gamma(A/H\rightarrow \nu\bar{\nu})&=&\frac{\kappa_{\nu,1}^2+\kappa_{\nu,2}^2+\kappa_{\nu,3}^2}{32\pi}M_{H_2}\nonumber\\
&\approx&\begin{cases}\displaystyle 5.5\times10^{-4}\left[1+1.0\left(\frac{m_1}{0.03~{\rm eV}}\right)\right]\left(\frac{M_{H_2}}{5~{\rm TeV}}\right)^4\left(\frac{\kappa_s}{0.2}\right)^{-2}\,M_{H_2}\quad&{\rm Normal~hierarchy}\\
\displaystyle1.0\times10^{-3}\left[1+0.16\left(\frac{m_3}{0.016~{\rm eV}}\right)\right]\left(\frac{M_{H_2}}{5~{\rm TeV}}\right)^4\left(\frac{\kappa_s}{0.2}\right)^{-2}\,M_{H_2}&{\rm Inverted~hierarchy}\end{cases}\label{eq:neut_neu}
\end{eqnarray}
\end{widetext}

\begin{figure*}[tb]
\begin{center}
\subfigure[]{\includegraphics[width=0.45\textwidth,clip]{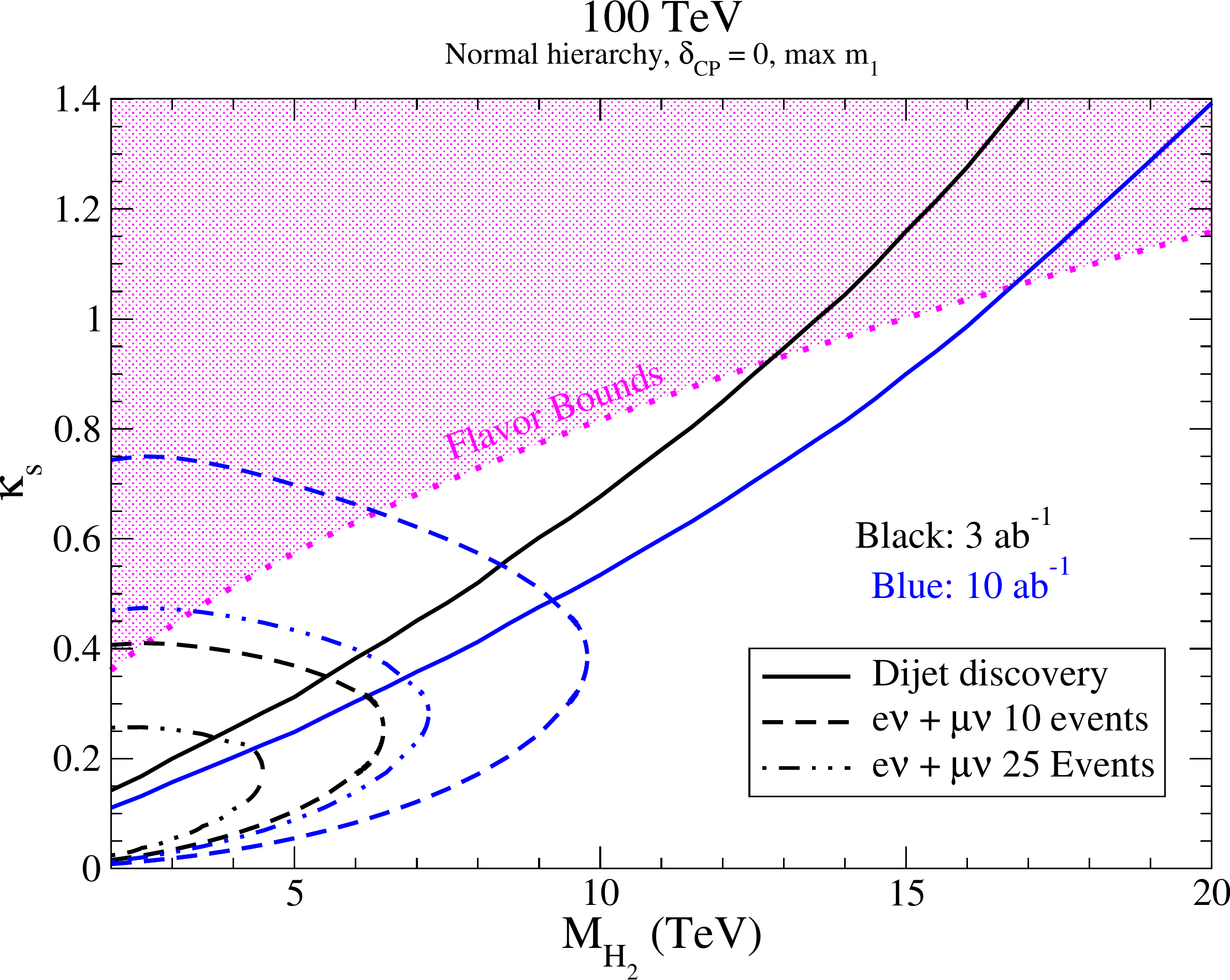}\label{fig:discovery_a}}
\subfigure[]{\includegraphics[width=0.45\textwidth,clip]{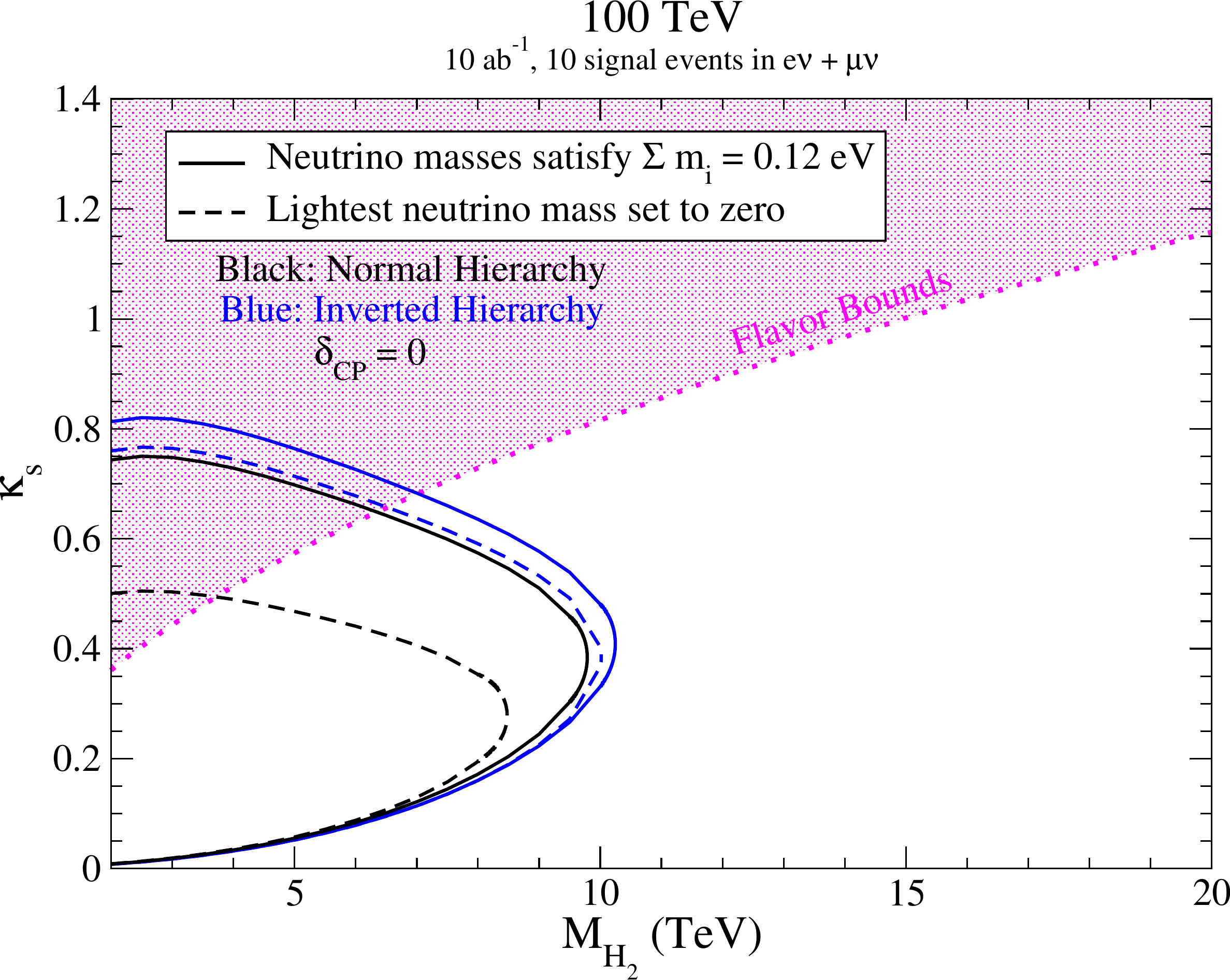}\label{fig:discovery_b}}
\caption{\label{fig:discovery} (a) {\color{black}$5\sigma$ discovery} reaches of $\kappa_s$ at a 100 TeV pp collider as a function of the $H_2$ mass with (black) 3 and (blue) 10 ab$^{-1}$ of data.  The normal hierarchy is used, $\delta_{CP}=0$, and the lightest neutrino mass is set to its maximum value.  Dijet discovery is possible for parameter values above the solid lines, and within the dashed (dot-dashed) curve 10 (25) events can be observed in $pp\rightarrow H^\pm\rightarrow e\nu+\mu\nu$. The shaded region is ruled out by current flavor constraints.  (b) Regions of parameters space needed to observe 10 events with 10 ab$^{-1}$ in the $e\nu+\mu\nu$ channel for (black) the normal hierarchy and (blue) the inverted hierarchy.  The region inside the solid lines are for when neutrino masses satisfy $\sum m_i=0.12$~eV and dashed lines for when the lightest neutrino is massless.  We have set $\delta_{CP}=0$. Dijet projections are the same as in panel (a).}
\end{center}
\end{figure*}   

As discussed previously, the major production mode of the heavy Higgs bosons is via $s$-channel production.  Pair production through EW gauge bosons only depends on gauge couplings and is therefore more model independent than $s$-channel production.  However, the pair production rates are exceedingly small.  At a 14 TeV LHC, pair production for two 2 TeV scalars is $\sim 0.05$ ab while at 100 TeV the pair production rate is $\sim 0.1$ fb~\cite{Davoudiasl:2019lcg}.  For $\kappa_s=0.2$, the single production of a 2 TeV heavy scalar boson is $\sim 4-8$~fb at the 14 TeV LHC and {\color{black}$\sim 0.9$} pb at 100 TeV\footnote{{\color{black}The range of cross sections denote the difference between charged and neutral scalar production.  Due to the high $x$ pdfs, at the LHC this difference is more pronounced.  At 100 TeV, these rates are within $10\%$ of each other.}}.  Hence, we focus on single production.  Additionally, a $4$ fb cross section can create $\sim 10,000$ heavy Higgs states with 3 ab$^{-1}$ of data at the LHC.  {\color{black}There are two major signal modes: di-jet with $H/A/H^\pm \rightarrow jj$ and lepton plus missing energy with $H^\pm \rightarrow e\nu/\mu\nu$.  While this is a relatively large rate, the QCD di-jet backgrounds are large and branching ratios into lepton plus neutrino are percent level. In the di-jet channel,} extrapolating to 3 ab$^{-1}$ of data from current ATLAS~\cite{ATLAS:2019fgd} and CMS~\cite{CMS:2019gwf} results, this translates to {\color{black} ruling out $\kappa_s\gtrsim 0.3-0.6$} at the $2\sigma$-level.\footnote{{\color{black}The range of $\kappa_s$ is due to ATLAS and CMS having different exclusion regions.  For details on the extrapolation method, see Ref.~\cite{Davoudiasl:2021syn}}.}  However, as seen in Fig.~\ref{fig:Dmixing}, these values are already ruled out by flavor constraints.   As the mass increases, the HL-LHC di-jet exclusion become significantly worse than the flavor constraints.    {\color{black} In the lepton plus neutrino channels, for a 2 TeV charged scalar with $\kappa_s=0.2$, the $pp\rightarrow H^\pm \rightarrow e\nu/\mu\nu$ cross sections are $\sim (3-7)\times 10^{-5}$ pb, where this range encompasses both hierarchies and electrons/muons.  Using a simple root luminosity extrapolation of current ATLAS~\cite{ATLAS:2019lsy} and CMS~\cite{CMS-PAS-EXO-19-017} searches for $W'\rightarrow e\nu/\mu\nu$, we find with 3 ab$^{-1}$ of data the experiments could be sensitive to cross sections of $(8-17)\times 10^{-5}$ pb.  This is tantalizingly close to our signal cross section, with the inverted hierarchy in the $H^\pm\rightarrow e\nu$ mode being the most promising signal to observe.  Although this may bear more detailed studies, we have assumed that both statistical and systematic uncertainties scale with luminosity.  The true sensitivity will depend on the details of improvements in systematic uncertainties in these channels.}  Hence, we focus on signals at a 100 TeV pp collider.

In Fig.~\ref{fig:discovery} we show the discovery reach as a function of the heavy Higgs mass at a 100 TeV pp collider with 3 and 10 ab$^{-1}$ of data.  Both (a) and (b) consider single production for $H^\pm,H,A$.  Single production cross sections are generated via \texttt{MadGraph5\_aMC@NLO}~\cite{Alwall:2014hca} with the model implemented using \texttt{FeynRules}~\cite{Christensen:2008py,Alloul:2013bka}.   We consider all four heavy states, $H^\pm,H,A$, to have degenerate masses and add their cross sections incoherently.  Their collective mass is labelled as $M_{H_2}$.  Indeed, this assumption greatly alleviates the bounds from the oblique parameters~\cite{Peskin:1990zt,Peskin:1991sw,Barbieri:2006dq,Haber:2010bw,Ahriche:2015mea}.

The discovery reach in the normal hierarchy is shown in Fig.~\ref{fig:discovery_a}.  We set $\delta_{CP}=0$ and $m_1$ to the maximum that saturates the cosmological bound $\sum m_i\lesssim 0.12$~eV.  Areas above the solid curves could be discovered in the dijet channel, where we have adapted the projected dijet search reach~\cite{Golling:2016gvc}\footnote{See Ref.~\cite{Davoudiasl:2021syn} for details on how we project dijet searches.}.    Areas within the dotted and dash-dot curves can be discovered with 10 or 25 total events, respectively, in the $pp\rightarrow H^\pm\rightarrow e\nu$ and $pp\rightarrow H^\pm \rightarrow\mu\nu$ channels.   {\color{black}The number of events for discovery is chosen considering zero background, we need 25 events to discover with Gaussian statistics and $\sim 10$ with Poisson (assuming 1 background event).  A more complete study with a full background simulation would be needed to determine the precise discovery region.  However, we expect our estimates to encompass the true range since background will be small.}

{\color{black}We can understand the curves in~\ref{fig:discovery_a}} by noting that $\kappa_s$ governs the production cross section.  Hence, at low $\kappa_s$ not enough $H^\pm$ are produced to create a signal.  At high $\kappa_s$ there is enough cross section, but decays into jets dominate, and the branching ratios ${\rm BR}(H^\pm \rightarrow e\nu)$ and ${\rm BR}(H^\pm \rightarrow \mu\nu)$ are too small to generate enough events.  Hence, the leptonic searches enclose an area in the $\kappa_s-M_{H_2}$ instead of providing the simple lower bound of the dijet searches.  The regions ruled out by flavor searches discussed earlier.

In Fig.~\ref{fig:discovery_b}, the regions enclosed by the curves are the parameter regions that will provide 10 events in the $e\nu$ and $\mu\nu$ channels with 10 ab$^{-1}$ of data.  We show the regions for the normal and inverted hierarchies, with the respective lightest neutrino masses set to zero or saturating the cosmological bound $\sum m_i\lesssim 0.12$~eV.  As can be seen, the 100 TeV collider is always more sensitive to the inverted hierarchy.  This can be understood by noting that Fig.~\ref{fig:BRs} and Eqs.~(\ref{eq:norm_el}-\ref{eq:norm_tau},\ref{eq:inv_el}-\ref{eq:inv_tau}) clearly indicate that in both hierarchies we obtain similar amounts of muons, but the inverted hierarchy produces many more electrons.  

The dijet bounds depend very weakly on the hierarchy.  For large $\kappa_s$ this is unsurprising since dijet decays dominate the total width, and the branching ratio into dijets will depend very little on the details of the neutrino hierarchy.  However, for small $\kappa_s$ the partial widths into neutrinos can dominate the total width.  As Eqs.~(\ref{eq:norm_el}-\ref{eq:norm_tau},\ref{eq:inv_el}-\ref{eq:inv_tau},\ref{eq:neut_neu}) indicate, the widths into neutrinos depend strongly on the neutrino hierarchy.  In this case, the production cross section times branching ratio into jets is schematically
\begin{eqnarray}
\sigma(pp\rightarrow H^\pm/H/A){\rm BR}(H^\pm/H/A\rightarrow jj)\sim \frac{\kappa_s^4}{\Gamma_\nu},
\end{eqnarray}
where $\Gamma_\nu$ is a generic partial width into neutrino final states.  At a fixed mass, there is an upper bound on this cross section times branching ratio.  However, even a factor of two change in $\Gamma_\nu$ can be compensated by a $2^{1/4}\approx 1.2$ change in $\kappa_s$.  That is, a large change in the branching fraction into neutrinos can be compensated by a small $\mathcal{O}(10\%)$ change in $\kappa_s$.  In the region where this important, $\kappa_s\ll 1$, this change is not visible on the plots.

\section{Summary and Conclusions}

In this work, we considered the possibility that the small neutrino masses deduced from flavor oscillation experiments originate from the QCD quark condensate.  We focused on the case of Dirac neutrinos, where the requisite effective interaction is mediated by a dimension-6 operator.  The required coupling to light quarks makes hadron colliders natural settings for investigation of the underlying model.  We showed that charged meson decay constraints can be avoided if strange quarks provide the dominant effect.  This scenario can be realized with one additional Higgs doublet at the TeV scale, giving rise to the dimension-6 interactions of quarks, leptons, and right-handed neutrinos.  We described how the structure of the heavy doublet Yukawa couplings may be realized by appropriate UV model building.  

Flavor data from $D -\bar D$ meson mixing constrain the allowed parameter space of the model.  However, much of the remaining parameter space can be tested through resonant production of heavy Higgs states from $s\bar s$ and $c\bar{s}/s\bar{c}$ initial states at a future 100 TeV hadron collider.  Both dijet and lepton + missing energy final states can in general be accessible, for generic choices of parameters.  An interesting and uncommon aspect of our model is that measurements of the heavy charged Higgs branching ratios into the lepton flavor final states can distinguish between the underlying ``normal" and ``inverted" neutrino mass hierarchies.  Specifically, in our model leptonic charged Higgs decays are dominated by muons in the case of normal hierarchy, followed by taus and electrons; for inverted hierarchy the charged lepton ordering is reversed.  Thus, collider probes of our scenario can be complementary to corresponding tests of the mass ordering at future neutrino oscillation experiments, and shed light on the microscopic neutrino mass generation mechanism.

\begin{acknowledgments}
{\bf Acknowledgments:} The work of H.D. and M.S. is supported by the United States Department of Energy under Grant Contract DE-SC0012704.  I.M.L. is supported in part by the United States Department of Energy grant number DE-SC0017988.  The data to reproduce the plots are available upon request.
\end{acknowledgments}



\bibliography{qcd-numass-refs}

\end{document}